\documentclass[twocolumn,prl]{revtex4-1}
\usepackage[T1]{fontenc}
\usepackage{epsfig}
\usepackage{wasysym}
\usepackage{natbib}

\begin{document}
\title{WIMP diffusion in the Solar System including solar WIMP-nucleon scattering}
\author{Sofia Sivertsson}
\email{sofiasi@kth.se}
\affiliation{Department of Theoretical Physics, Royal Institute of Technology (KTH), AlbaNova University Center, SE-106 91 Stockholm, Sweden and\\
The Oskar Klein Centre for Cosmoparticle Physics, Stockholm University, AlbaNova University Center, SE-106 91 Stockholm, Sweden}
\author{Joakim Edsj\"o}
\email{edsjo@fysik.su.se}
\affiliation{Department of Physics \& The Oskar Klein Centre for Cosmoparticle Physics, Stockholm University, AlbaNova University Center, SE-106 91 Stockholm, Sweden}
\date{}
\begin{abstract}
Dark matter in the form of Weakly Interacting Massive Particles (WIMPs) can be captured by the Sun and the Earth, sink to their cores, annihilate and produce neutrinos that can be searched for with neutrino telescopes. The calculation of the capture rates of WIMPs in the Sun and especially the Earth are affected by large uncertainties coming mainly from effects of the planets in the Solar System, reducing the capture rates by up to an order of magnitude (or even more in some cases). We show that the WIMPs captured by weak scatterings in the Sun also constitute an important bound WIMP population in the Solar System. Taking this population and its interplay with the population bound through gravitational diffusion into account cancel the planetary effects on the capture rates, 
and the capture essentially proceeds as if the Sun and the Earth were free in the galactic halo. The neutrino signals from the Sun and the Earth are thus significantly higher than claimed in the scenarios with reduced capture rates. 
\end{abstract}
\date{July 10, 2012}
\maketitle

\section{Introduction}

The identity of dark matter is one of the major problems in physics. If WIMPs (weakly interacting massive particles) exist they naturally produce the correct relic abundance, making them one of the best dark matter candidates. There are many ongoing searches hoping to observe the particle nature of dark matter (see \cite{Bertone:2010zz}) for a review). Searches for neutrinos from WIMP annihilations in the Sun and the Earth are receiving a lot of interest with the recent completion of the IceCube neutrino telescope and the recently proposed dark disc component of the Milky Way dark matter halo \cite{Read:2008fh,Bruch:2009rp}.

The idea to search for signals from dark matter accumulating in Solar System objects by WIMP-nucleon scatterings, so called \emph{weak capture}, goes back to \cite{Press-Spergel,*Krauss:1985ApJ,*Freese:1985qw,*Krauss:1985aaa,*Gaisser:1986ha}. Even if the main ideas were laid out already in the mid 1980s, consensus has still not been reached for the capture rate of WIMPs in the Sun and the Earth. The main complication is that the Sun and the Earth are both part of the Solar System. 

For the Earth, things are complicated by the Earth being deep within the potential well of the Sun, accelerating the galactic WIMPs reaching the Earth. The history of determining the capture rate has gone through a series of rather abrupt changes as our understanding of the problem has evolved (for a review, see \cite{Lundberg}). For resonances, i.e.~when the WIMP mass matches the mass of an element in the Earth, the Earth can efficiently capture WIMPs directly from the galactic halo as in \cite{Gould:capture}. It was later realized that far from the resonances, i.e.~for heavy WIMPs, the Earth's low escape velocity only allows it to capture WIMPs of velocities lower than the solar escape velocity at the Earth's distance $r_{\oplus}$, i.e.\ only WIMPs already bound to the Solar System. It was however found in \cite{Gould:diffusion} that gravitational interactions with the planets will cause WIMPs to diffuse into the Solar System, thus forming a bound population of WIMPs. The gravitational interaction by the planets diffuse WIMPs between the population bound to the Solar System and the galactic WIMP population; given enough time this will result in an equilibrium configuration of WIMPs in the Solar System, referred to as \emph{gravitational equilibrium}. The gravitational equilibrium configuration is easily found since the gravitational interactions causing the diffusion obey Liouville's theorem, assuring that the diffusion preserves the phase space density. Hence, any WIMP phase space over- or under-density in the Solar System will, given enough time, be mixed with the large galactic WIMP reservoir and recover the equilibrium configuration, i.e.\ the same phase space density of Solar System bound WIMPs as for the galactic WIMP halo \cite{Gould:diffusion}. One could also argue for this from detailed balance. Since the diffusion in and out of the Solar System is symmetric, we will at some point reach an equilibrium between particles diffusing into and out of the Solar System, and this equilibrium will happen when the phase space densities are the same.

It was later realized \cite{GouldAlam} that the bound WIMP population could be reduced via \emph{solar depletion}, which occurs if WIMPs in the bound population are perturbed to end up in the \emph{solar loss cone}, i.e.\ on orbits crossing the Sun, allowing them to be captured by the Sun. If this process is more efficient than the gravitational diffusion refilling the bound population, a reduction could occur. This was studied with numerical simulations in \cite{Lundberg} and a reduction of the capture rate for heavy WIMPs of up to an order of magnitude was found (shown as the magenta, dotted curve in Fig.~\ref{fig:capture}).

We also mention a bound WIMP population proposed in \cite{Damour} emerging from weakly captured WIMPs whose orbits only pass the outskirts of the Sun. These WIMPs can be gravitationally disturbed by the planets onto orbits no longer crossing the Sun. It was argued in \cite{Damour} that this WIMP population will over the lifetime of the Sun build up a phase space density much higher than that of the galactic halo, making it important for WIMP capture in the Earth. Such gravitational interactions must obey Liouville's theorem, making it impossible for the proposed gravitational diffusion to create a phase space density larger than that of the solar loss cone. The latter will be discussed in detail later, for now we just conclude that the high phase space density of \cite{Damour} cannot exist. This population was also numerically investigated in \cite{Peter:first}, finding no support for a large enhancement.

For the Sun, the capture rate calculations are simpler as the Sun gravitationally dominates the Solar System. However, the WIMPs in the solar loss cone can be disturbed gravitationally by the planets, and it was shown in \cite{Peter:sun} that the WIMPs reaching out to Jupiter will be thrown out by Jupiter before having time to scatter again in the Sun, introducing a \emph{Jupiter depletion}. For WIMPs being heavy enough compared to the target nucleus, the energy loss in the scatter is small enough for the first bound orbit of all weakly captured WIMPs to reach $r_{\jupiter}$ and hence be perturbed before having time to scatter again in the Sun. The scattering in the Sun occurs either via spin-dependent scattering (mainly on hydrogen) or spin-independent scattering (also on heavier elements). In \cite{Peter:sun} it was concluded that Jupiter depletion substantially reduces the solar capture rate for WIMPs heavier than a TeV, which scatter predominantly spin-dependently in the Sun.

In this paper, we will show that Liouville's theorem is not only valid for gravitational diffusion but also holds approximately for weak capture. To a first approximation the phase space density in the solar loss cone is then the same as that of the gravitationally captured WIMPs.
Hence, weak capture should rather be viewed as another source of WIMP diffusion in the Solar System. 
For scattering cross sections that are not too low, the equilibrium time scale for the solar loss cone is shorter than the age of the Solar System. For a 100 GeV WIMP, equilibrium has been reached for spin-dependent scattering cross sections  (on protons) $\sigma_{\mathrm{SD}}\gtrsim 10^{-44}$ cm$^2$ or spin-independent scattering cross sections (on protons) $\sigma_{\mathrm{SI}}\gtrsim 10^{-46}$ cm$^2$. For heavier WIMPs, slightly larger cross sections are needed \cite{Sivertsson,Peter:sun}. However, not all regions of bound phase space have reached gravitational equilibrium, giving a region of lower phase space density, a hole, as shown in Fig.~3 in \cite{Gould:diffusion}. This so called \emph{Gould hole} affects velocities (with respect to the Earth) between $\sim 27$ km/s and $\sim 69$ km/s. Weak capture will populate a vertical strip at $\sim 30$ km/s in this hole.

The discussed solar and Jupiter depletions are both, in principle, gravitational diffusion between the two bound populations; as their phase space densities are approximately the same the effects, to a first approximation, cancel completely as the net flow between them vanishes. The studies of \cite{Peter:sun,Lundberg} still found these effects important since they only looked at diffusion in one direction, neglecting the reverse process. In \cite{Lundberg} they assumed the Sun to be optically thick to WIMPs, i.e.\ removed all WIMPs entering the solar loss cone, neglecting diffusion out of the solar loss cone. In the Jupiter depletion calculation of \cite{Peter:sun} the possibility of an inverse process, i.e.\ diffusion into the solar loss cone, was not considered. In the subsequent paper \cite{Peter:earth}, diffusion of WIMPs by Jupiter into the solar loss cone was actually observed but not investigated further.

%%%%%%%%%%
\section{Liouville's theorem and WIMPs in the solar loss cone}
For Liouville's theorem to be applicable, the force, $\mathbf{F}$, encountered by the particles must be differentiable (i.e.\ ``smooth'') and must fulfill $\nabla_p\cdot \mathbf{F}=0$ where $\nabla_p$ is the gradient operator in momentum space. Hence, Liouville's theorem is applicable to forces that do not depend on momentum.
Obviously, weak scattering satisfies neither of these requirements. However, we will see that weak capture can be well approximated by a force fulfilling Liouville's theorem.

WIMPs passing through the Sun typically scatter 0 or 1 times per passage. In such a scatter the WIMP loses the energy (over WIMP mass) $\Delta E$ uniformly distributed in the range
\begin{equation}
0\leq\Delta E\leq\frac{4Mm}{(M+m)^2}E_k,
\label{energy loss}
\end{equation}
where $M$ and $m$ are the masses of the WIMP and the target nucleus, respectively, and $E_k$ is the kinetic energy over WIMP mass of the WIMP just before the scatter. The WIMP scatter probability in the solar passage depends very weakly on the WIMP momentum, but the energy loss in the scatter goes as $\Delta E \propto E_k$, as seen in Eq.~(\ref{energy loss}). To fulfill Liouville's theorem the WIMP energy loss should rather be proportional to the time the WIMP spends inside the Sun, i.e.\ $\Delta E \propto E_k^{-1/2}$.

Since we ultimately want to investigate gravitational effects of the planets, we are only interested in WIMPs in the galactic halo or in the solar loss cone on orbits reaching out to the planets. The innermost planet in the Solar System is Mercury, at radius $r_{\mercury}$. A bound solar crossing WIMP on an orbit stretching out to at least $r_{\mercury}$ will at radius $r$ inside the Sun have kinetic energy in the range
\begin{equation}
\Psi_\odot(r)\left(1-\frac{1}{k}\frac{R_\odot}{r_{\mercury}}\right)\leq E_k\leq\Psi_\odot(r) \mbox{, \ \ } k\equiv\frac{\Psi_\odot(r)}{\Psi_\odot(R_\odot)}\geq 1, 
\label{Ek-range}
\end{equation}
where $R_\odot$ is the radius of the Sun and $\Psi_\odot(r)$ is the absolute value of the solar gravitational potential at the distance from the Sun $r$. As $R_\odot\simeq 0.012r_{\mercury}$, the WIMP kinetic energy inside the Sun is well approximated by $E_k\simeq\Psi_\odot(r)$, especially for orbits going further out to the more massive and more interesting planets. Furthermore, the typical WIMP velocities in the galactic halo are significantly lower than the escape velocity inside the Sun and unless we are on a resonance, only the low velocity WIMPs of the halo are captureable by the Sun, making $E_k\simeq\Psi_\odot(r)$ a good approximation also here. Since all WIMPs of interest to us have, essentially, the same velocity inside the Sun, the momentum dependence in the scatter is not important for the WIMPs of interest except, possibly, if we are close to a resonance.

For an outside observer, who does not notice the small momentum dependence, the random energy loss of WIMPs passing the Sun could be equally well-attributed to a time dependent deepening potential well inside the Sun, i.e.\ a differentiable force fulfilling Liouville's theorem. Hence we conclude that Liouville's theorem can be applied to high precision for the weakly captured WIMPs that are subject to gravitational interaction with the planets. 

In the language of detailed balance, we can see this approximate relationship from the fact that the rate at which WIMPs scatter into the solar loss cone is approximately the same as the rate at which they leave via further scatterings in the Sun (as the velocity dependence is so weak). This means that the phase space density of WIMPs in the solar loss cone will be approximately the same as in the galactic halo. As we move close to the Sun (within the radius of Mercury), the approximation of equal rates worsens (Liouville's theorem is less applicable), and the density will be different. We are, however, not concerned with those orbits.

We will now verify the above reasoning by calculating the density in the solar loss cone using Liouville's theorem and compare with the numerical results of \cite{Sivertsson}. 
We assume a Maxwell-Boltzmann distribution for the galactic halo as in \cite{Sivertsson} but this will not be important for our conclusions. The same arguments will hold also for other velocity distributions.
The relevant galactic phase space density is then, viewed by an observer (i.e.~the Sun) moving with the rotational velocity $v_\odot=\sqrt{2/3}\bar v$, given by \cite{Gould:capture}
\begin{equation}
F(\mathbf{u})=n_W\frac{1}{\mbox{e}\pi^{3/2}}\left(\frac{3}{2}\right)^{3/2}\frac{1}{\bar v^3}\exp\left(-\frac{3u^2}{2\bar v^2}\right)\frac{\sinh(\sqrt{6}u/\bar v)}{\sqrt{6}u/\bar v}.
\label{ph-dens}
\end{equation}
Here $u=|\mathbf{u}|$ and $n_W$ are the local, galactic WIMP velocity and number density far from the Sun, respectively; $\bar v=270$ km/s is the three-dimensional velocity dispersion. $F(\mathbf{u})$ maximizes as $u\to 0$ and is approximately constant for low velocities, i.e.~for $u\ll \bar v$. Since only the very low velocity WIMPs in the galactic halo are prone to planetary gravitational diffusion, the gravitational equilibrium phase space density for WIMPs in the Solar System is $F(\mathbf{0})$ \cite{Gould:diffusion}, which is also the equilibrium configuration with maximal density. The maximal phase space density in the solar loss cone is then also $F(\mathbf{0})$ and is achieved
if weak capture only probes very low $u$, i.e.~for $M\gg m$ as seen in Eq.~(\ref{energy loss}). 

\begin{figure}
\centerline{
\includegraphics[width=0.49\textwidth]{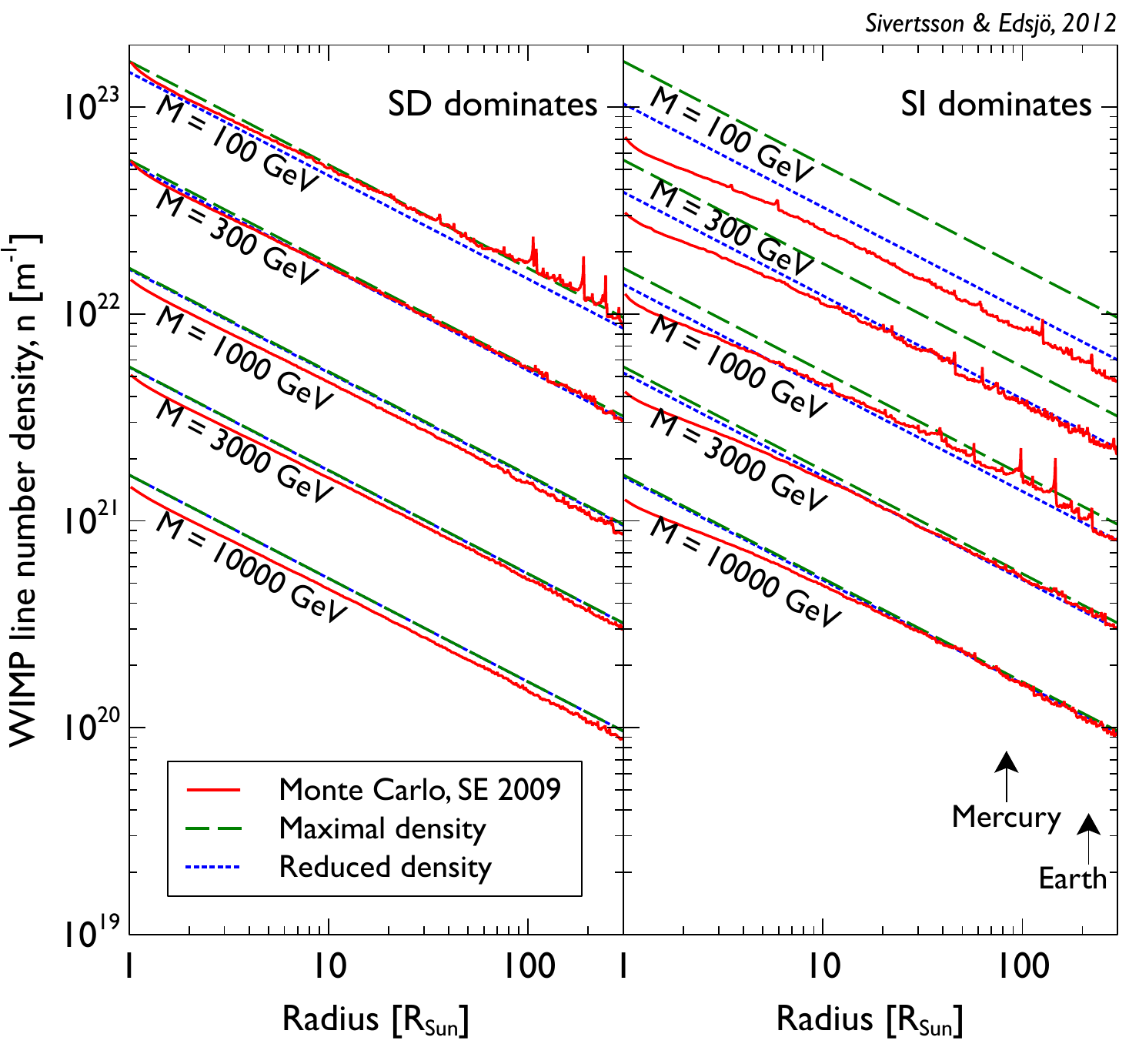}
}
\caption {The WIMP line number density $n(r)=4\pi r^2 n_W^{\odot}(r)$ in the solar loss cone for different WIMP masses with spin-dependent scatter dominating in the Sun (left figure) and spin-independent scatter dominating (right figure). The green, long-dashed lines are the maximal densities by Liouville's theorem, i.e.\ for $M\gg m$, as in Eq.~(\ref{suncrossnumdens}). The blue, dotted lines show the density by Liouville's theorem but taking into account that $F(\mathbf{u})$ is not flat and is hence valid also for $M\sim m$. These are compared with the numerical results from SE 2009 \cite{Sivertsson} shown with red, solid lines.}
\label{fig:numdens}
\end{figure}

Knowing the phase space density, the spatial density in the solar loss cone is simply the integral over the allowed angular momentum for velocities below the local escape velocity $v_e(r)$ with $r$ being the distance from the solar centre. For a bound WIMP's orbit to cross the Sun, the angular momentum over WIMP mass $J$ must fulfill $J\leq J_{\max}=wR_\odot$, where $w$ is the local velocity at $R_\odot$. As for $E_k$ in Eq.~(\ref{Ek-range}), $J_{\max}$ is essentially the same for all WIMPs of interest, giving
$J_{\max}\simeq \sqrt{2GM_\odot R_\odot}$.
One can show that for an isotropic velocity distribution the probability for a bound WIMP at radius $r$ with velocity $u$ to be on a solar-crossing orbit, i.e.~to fulfill $J\leq J_{\max}$, is
$P_\odot(u,r)=1-\sqrt{1-(J_{\max}/ru)^2}$ for $ru>J_{\max}$
and $P_\odot=1$ otherwise. The maximal number density of WIMPs in the solar loss cone from Liouviulle's theorem is then
\begin{eqnarray}
n_W^\odot(r)&=&\int_0^{v_e(r)}4\pi u^2F(\mathbf{0})P_\odot(u,r)\mbox{ d}u\nonumber \\
&\simeq& n_\chi\frac{6\sqrt{3}}{e\sqrt{\pi}}\frac{1}{\bar v^3}\left(\frac{GM_\odot}{R_\odot}\right)^{3/2}\left(\frac{r}{R_\odot}\right)^{-5/2},
\label{suncrossnumdens}
\end{eqnarray}
where we have made a Taylor expansion in $R_\odot/r$ which gives an error $< 0.3\%$ for $r\geq r_{\mercury}$.
This result is shown in Fig.~\ref{fig:numdens} (green, dashed curve) and compared with the detailed numerical calculation in \cite{Sivertsson}. They fit very well when the assumption $M\gg m$ is fulfilled (which is always the case for spin-dependent scattering). The numerical simulations in \cite{Sivertsson} makes no assumptions on Liouville's theorem and consists of a Monte Carlo study following WIMP orbits to build up the bound population. That the agreement with our theoretical calculations is so good verifies that Liouville's theorem indeed applies here. Note that this figure is independent of the magnitude of the scattering cross sections. (It does depend on if capture is dominated by spin-independent or spin-dependent scatterings though.) See \cite{Sivertsson} for more details.

If $M\sim m$, the Sun also captures galactic halo WIMPs of higher velocities, which renders $F(\mathbf{u})\simeq F(\mathbf{0})$ (i.e.~$u\ll\bar v$) a bad approximation. Taking this into account reduces the normalization of the solar loss cone density with the factor $F(\mathbf{u})/F(\mathbf{0})$ averaged over the galactic velocities $u$ of the captured WIMPs (which reach the planets). We have performed this calculation numerically and the result is shown in Fig.~\ref{fig:numdens} (blue, dotted line), giving a better fit to the numerical results for $M\sim m$.
We see that our prediction fits very well with the numerical results with a slightly worse fit close to the iron resonance, i.e.\ for spin-independent scattering with $M=100$ GeV. We believe that this is due to Liouville's theorem not being a good approximation for the first scatter so close to the resonance since then all halo velocities of WIMPs bound to the Milky Way are prone to weak capture, making $E_k\simeq\Psi_\odot(r)$ of Eq.~(\ref{Ek-range}) a less good approximation. Spin-independent capture of light WIMPs in the Sun occur mainly on helium, carbon, nitrogen, oxygen and iron. As the masses vary, there will always be some significant capture not close to a resonance, the error in the blue curve in Fig.~\ref{fig:numdens} is thus not expected to be much larger for lighter WIMPs. Also, the bound population is not very important close to the resonances (see Fig.~\ref{fig:capture}), hence the discrepancy is not very important. For smaller radii than the planets, the agreement in Fig.~\ref{fig:numdens} is worse, but this is expected as several of our approximations break down in the limit $r \sim R_\odot$; this is of no concern to us as we are only interested in the WIMP density at $r > r_{\mercury}$.

%%%%%%%%%%
\section{Earth's WIMP capture rate and solar depletion}

\begin{figure*}
\centerline{
\includegraphics[width=0.49\textwidth]{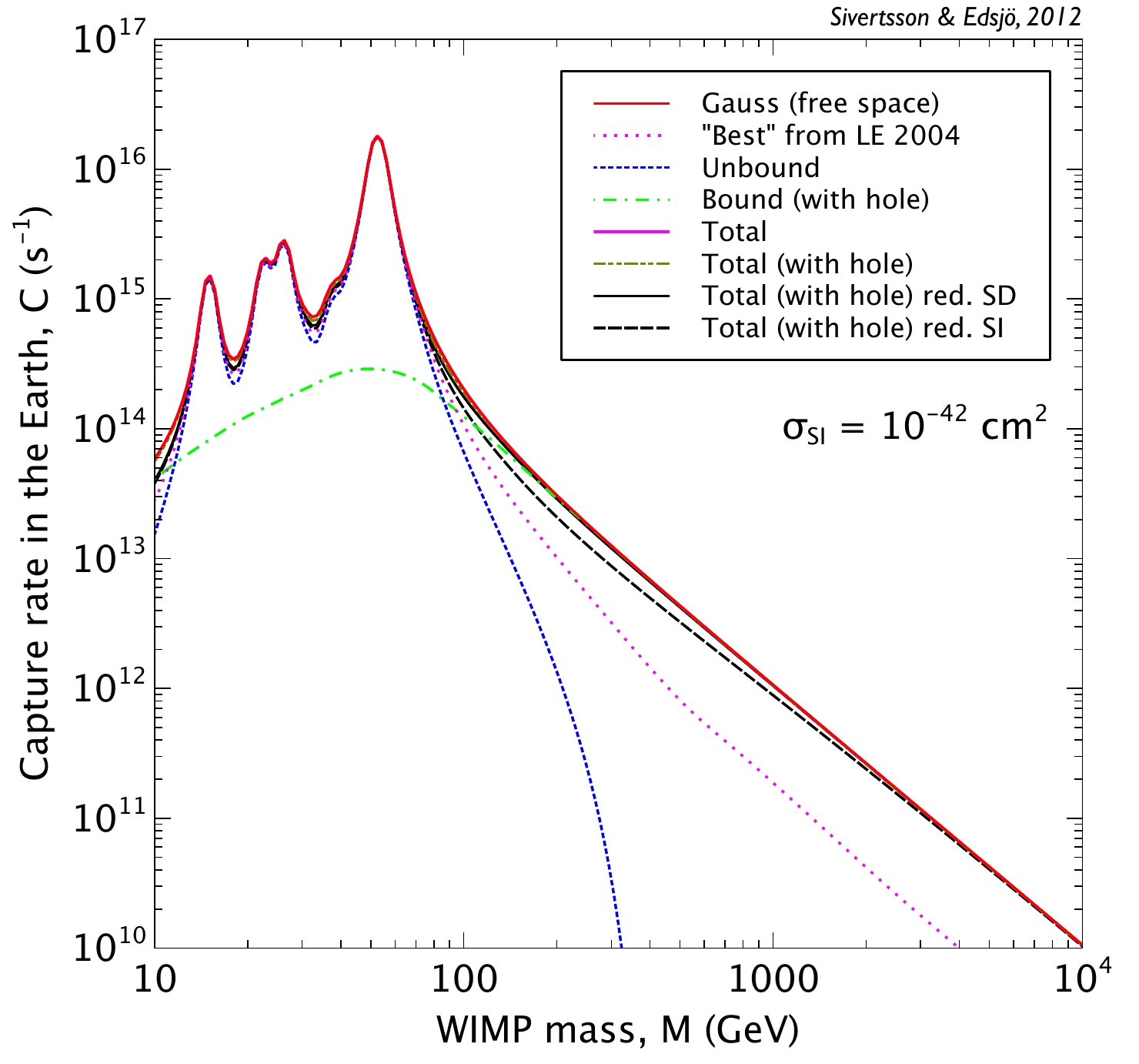}
\includegraphics[width=0.49\textwidth]{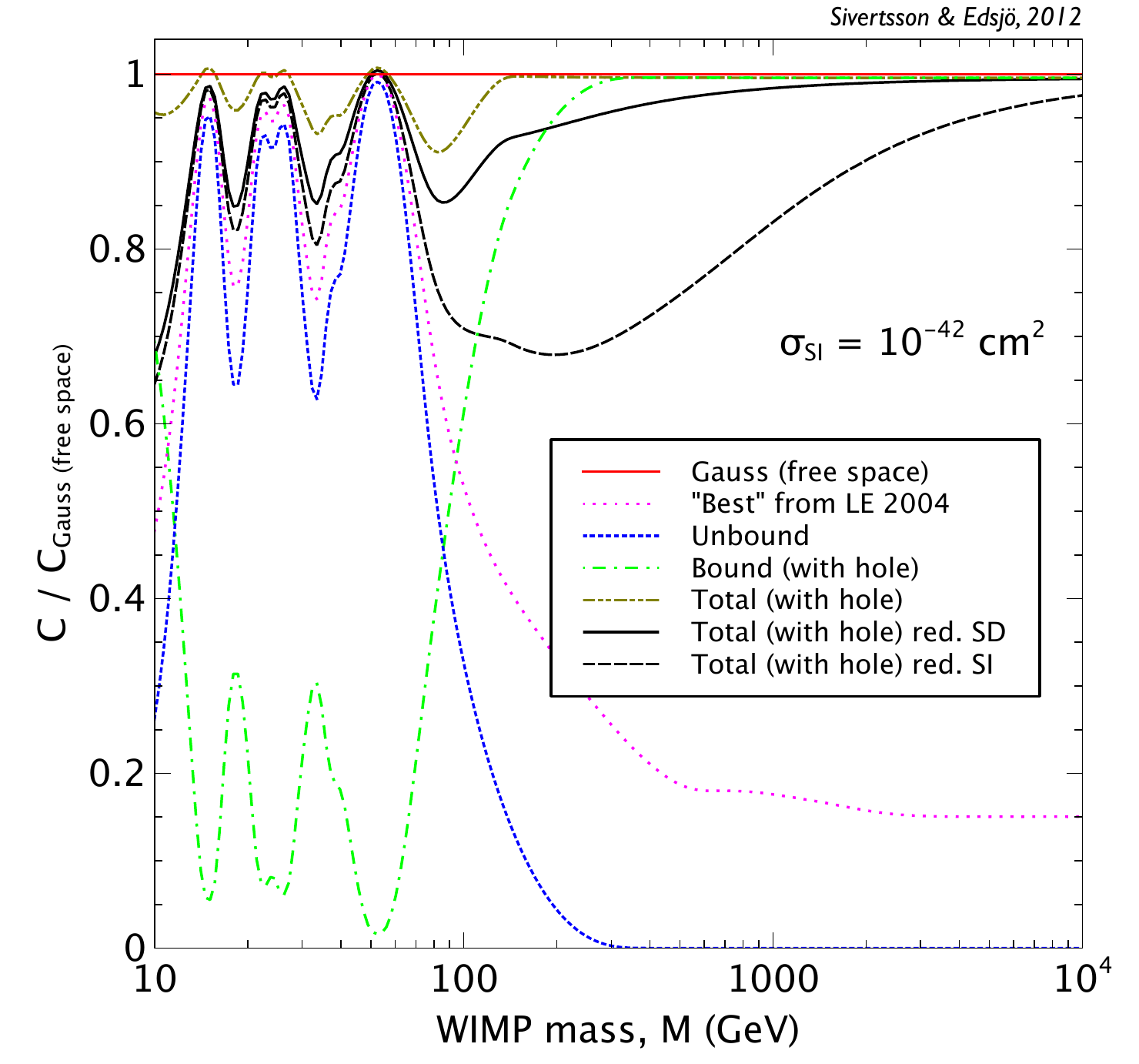}
}
\caption {The capture rate in the Earth for different phase space densities. We show the Gaussian (Maxwell-Boltzmann) at the Sun's location but for the Earth as in free space, the ``best'' curve from \cite{Lundberg}, the unbound population, the bound population with Gould's hole empty, the total with the hole and the total with the maximal reduction of the bound densities for spin-dependent (SD) and spin-independent (SI) scattering respectively dominating in the Sun. In the left figure, we show the capture rates and in the right the ratios with respect to the free space Gaussian. The cross section chosen is just for reference and the capture rate scales linearly with it.}
\label{fig:capture}
\end{figure*}

\begin{figure}
\centerline{
\includegraphics[width=0.49\textwidth]{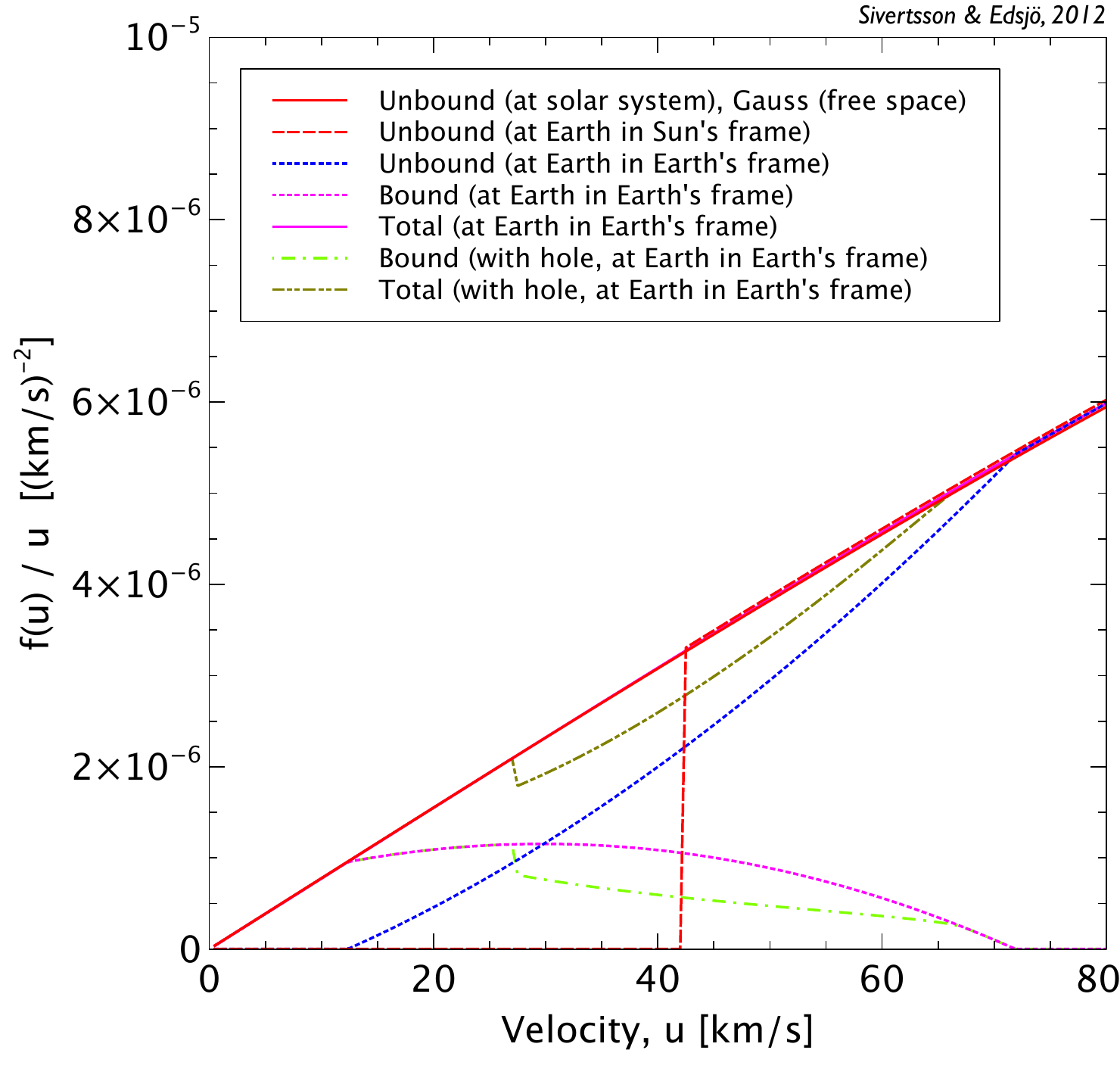}
}
\caption{The low velocity part of the velocity distributions (divided by velocity) of the various unbound and bound populations assuming a standard Maxwell-Boltzmann velocity distribution in the halo. Without inclusion of the Gould hole, the bound + unbound population almost perfectly matches the free space approximation. The Gould hole (i.e.\ taking into account that part of the phase space for the bound population is not filled), affect intermediate velocities and therefore has no dramatic effect on the capture rates shown in Fig.~\ref{fig:capture}}.
\label{fig:veldist}
\end{figure}

To make a conservative estimate on how much solar depletion can affect the Earth's capture rate, we assume the worst case scenario, i.e.\ that the refilling of the solar loss cone and the diffusion between the two populations are so efficient that the phase space density of all bound WIMPs is reduced to that of the solar loss cone. The capture rate for this conservative case is shown in Fig.~\ref{fig:capture}, concluding that even this maximal effect is indeed very small, at most a reduction by $\sim 35$\% for specific WIMP masses. The reduction is even smaller if spin-dependent scatter dominates in the Sun. The reason for the small reduction is that, for low enough WIMP masses to significantly reduce the solar loss cone density, the Earth
efficiently captures WIMPs directly from the galactic halo. As equilibrium between capture and annihilation has typically not been reached in the Earth, the annihilation rate 
goes as the square of the capture rate (see e.g.\ \cite{Lundberg}), making the boost of the annihilation rate even larger compared to \cite{Lundberg}.

In Fig.~\ref{fig:capture}, we also show the capture rate for the `standard' approximation \cite{Gould:diffusion} of taking the velocity distribution as if the Earth were in free space (i.e.\ at the location of the Solar System, but not affected by the gravitational attraction by the Sun). Also shown are the capture rates directly from the galactic halo, i.e.\ the unbound population, as well as from the population bound to the Solar System with the Gould hole in Fig.~3 in \cite{Gould:diffusion}. From Fig.~\ref{fig:capture}, we see that the capture rate is not very sensitive to solar depletion or the Gould hole, concluding that the free space approximation is actually quite good.

For clarity, we also show the velocity distributions for different WIMP populations that go into the capture rate calculations. We can define the one-dimensional velocity distribution $f(u)$ as
$
f(u) = \frac{4 \pi u^2}{n_w} F(\mathbf{u})
$
with $F(\mathbf{u})$ from Eq.~(\ref{ph-dens}). In the capture rate calculations, $f(u)$ enters as $f(u)/u$ (see e.g.\ \cite{Gould:diffusion,Lundberg} for details), which is shown in 
Fig.~\ref{fig:veldist}. We can clearly see that in the absence of the Gould hole, the bound plus unbound population almost perfectly matches the free space velocity distribution, as argued above. 
The Gould hole reduces the velocity distributions between $\sim 27$ km/s and $\sim 69$ km/s. The bound population is also affected by weak capture with the possible reduction discussed above.

We have assumed a Maxwell-Boltzmann velocity distribution, but our method applies to any velocity distribution. If the velocity distribution is very peaked at low velocities, the reduction in the solar loss cone could be larger, i.e.\ the difference between the green, dashed curves and blue, dotted curves in Fig.~\ref{fig:numdens} could be larger. Inversely, if the velocity distribution is peaked at high velocities (for a stream e.g.), we could get an enhancement instead of a reduction from weak capture.

%%%%%%%%%%
\section{Planetary effects on the Sun's capture rate}

Jupiter is with its mass of 318 Earth masses the planet in the Solar System with by-far the largest gravitational impact, as observed in \cite{Peter:sun} in removing WIMPs from the solar loss cone. The other consequence of this is that Jupiter gravitationally diffuses WIMPs efficiently enough to have achieved gravitational equilibrium for all parts of phase space representing bound orbits crossing $r_{\jupiter}$ \cite{Gould:diffusion,Lundberg}. Jupiter depletion is only important for $M\gg m$ but the solar loss cone phase space density is then the same as that of gravitational equilibrium. Hence, Jupiter by Liouville's theorem does not alter the solar loss cone density at $r_{\jupiter}$ since WIMPs are equally diffused into as out of the solar loss cone, leaving no net effect on the Sun's WIMP capture rate.

Another way to see this is to calculate the solar capture rate from $r_{\jupiter}$ instead of from $r\to\infty$. As gravitational equilibrium has been reached at $r_{\jupiter}$, the total phase space density (bound plus unbound WIMPs) is the same at $r_{\jupiter}$ as in the galactic halo, i.e.~$F(r_{\jupiter},\mathbf{u})=F(\infty,\mathbf{u})=F(\mathbf{u})$. The solar WIMP capture rate from $r_{\jupiter}$ is then the rate of WIMPs scattering in the Sun to orbits with $r_{\max}< r_{\jupiter}$, analogous to the standard scenario where the WIMPs from the galactic halo scatter to an orbit with $r_{\max}<\infty$. The deviation of the capture rate from $r_{\jupiter}$ compared to the standard scenario can be incorporated by assigning an ``effective'' escape velocity $\tilde{v}_e$ defined as the minimal velocity required at the location of the scatter to reach $r_{\jupiter}$. Comparing $\tilde{v}_e$ with the usual escape velocity $v_e$ gives
\begin{equation}
\frac{\tilde{v}_e}{v_e}=\sqrt{1-\frac{R_\odot}{k r_{\jupiter}}},
\label{jupiter-capture}
\end{equation} 
where $k\geq 1$ depends on the location of the scatter and is defined in Eq.~(\ref{Ek-range}). As $r_{\jupiter}\simeq 1119 R_{\odot}$ we get that $\tilde{v}_e \simeq v_e$, concluding that the Sun's capture rate is essentially unaffected by the gravitational presence of Jupiter. This presents an alternative way of reaching the same conclusion as we did before, i.e.\ that the influence of Jupiter alone is not important for the solar WIMP capture rate.

In this paper we have assumed the phase space density at $r_{\jupiter}$ to really be $F(\mathbf{0})$, as in \cite{Gould:diffusion}. This is supported by the numerical work in \cite{Lundberg}. In \cite{Peter:earth}, a slightly lower bound phase space density is found for some velocities. We argue that this is probably due to having neglected regions of longer equilibrium time scales (retrograde WIMPs with high angular momentum) as they are below the resolution limit of the simulations. It is argued in \cite{Peter:earth} that these WIMPs are not important by referring to Fig.~3 in \cite{Peter:earth}, but also there the resolution of the relevant WIMPs is too low to draw any conclusions on this population.

%%%%%%%%%%
\section{Summary and discussion}

We have shown that one can to high precision also use Liouville's theorem for weakly captured WIMPs, not just for the gravitationally captured WIMPs as previously believed. 
The solar weak capture process is hence to be viewed as an extra source of WIMP diffusion in the Solar System.

Both the effects of solar and Jupiter depletion have in the literature been found to be the most pronounced for heavy WIMPs, i.e.\ when the phase space density of the solar loss cone is the same as that of gravitational equilibrium. As both these effects are really gravitational diffusion between the solar loss cone and the population bound through gravitational diffusion, the effects are by Liouville's theorem cancelled by diffusion in the reverse direction. This results in the overall conclusion that WIMP capture in the Sun and the Earth can generally be treated as if they are both free in the galactic halo, i.e.\ returning to the conclusion of \cite{Gould:diffusion}. Only the Earth's capture rate can deviate from this with, at most, 30-35\% where the most important factor is the nature of weak capture in the Sun. Note that this is under the conservative assumption of weak capture being much faster than gravitational diffusion. In reality we expect a capture rate somewhere in between the conservative and the unreduced (total) capture rate (i.e.\ between the black curves and the dark green, dot-dot-dashed curve in Fig.~\ref{fig:capture}), depending on the relative efficiency of weak and gravitational diffusion.

The only way to achieve a higher phase space density of bound WIMPs in the Solar System is to either have a mechanism that violates Liouville's theorem, to have a higher WIMP phase space density in the galactic halo, or if the Solar System in some way was born with a dark matter overdensity (of which a small portion, e.g.\ in the Gould hole, could survive until today). The second case is achieved by the galactic dark matter having a disc structure which co-rotates with the stars, increasing the phase space density of the important low velocity WIMPs \cite{Read:2008fh,Bruch:2009rp}.

\section*{Acknowledgements}
We acknowledge support from the Swedish Research Council (Contract Nos.\ 621-2010-3301 and 315-2004- 6519). 
\bibliography{references}

\begin{thebibliography}{17}%
\makeatletter
\providecommand \@ifxundefined [1]{%
 \@ifx{#1\undefined}
}%
\providecommand \@ifnum [1]{%
 \ifnum #1\expandafter \@firstoftwo
 \else \expandafter \@secondoftwo
 \fi
}%
\providecommand \@ifx [1]{%
 \ifx #1\expandafter \@firstoftwo
 \else \expandafter \@secondoftwo
 \fi
}%
\providecommand \natexlab [1]{#1}%
\providecommand \enquote  [1]{``#1''}%
\providecommand \bibnamefont  [1]{#1}%
\providecommand \bibfnamefont [1]{#1}%
\providecommand \citenamefont [1]{#1}%
\providecommand \href@noop [0]{\@secondoftwo}%
\providecommand \href [0]{\begingroup \@sanitize@url \@href}%
\providecommand \@href[1]{\@@startlink{#1}\@@href}%
\providecommand \@@href[1]{\endgroup#1\@@endlink}%
\providecommand \@sanitize@url [0]{\catcode `\\12\catcode `\$12\catcode
  `\&12\catcode `\#12\catcode `\^12\catcode `\_12\catcode `\%12\relax}%
\providecommand \@@startlink[1]{}%
\providecommand \@@endlink[0]{}%
\providecommand \url  [0]{\begingroup\@sanitize@url \@url }%
\providecommand \@url [1]{\endgroup\@href {#1}{\urlprefix }}%
\providecommand \urlprefix  [0]{URL }%
\providecommand \Eprint [0]{\href }%
\providecommand \doibase [0]{http://dx.doi.org/}%
\providecommand \selectlanguage [0]{\@gobble}%
\providecommand \bibinfo  [0]{\@secondoftwo}%
\providecommand \bibfield  [0]{\@secondoftwo}%
\providecommand \translation [1]{[#1]}%
\providecommand \BibitemOpen [0]{}%
\providecommand \bibitemStop [0]{}%
\providecommand \bibitemNoStop [0]{.\EOS\space}%
\providecommand \EOS [0]{\spacefactor3000\relax}%
\providecommand \BibitemShut  [1]{\csname bibitem#1\endcsname}%
\let\auto@bib@innerbib\@empty
%</preamble>
\bibitem [{\citenamefont {Bertone}(2010)}]{Bertone:2010zz}%
  \BibitemOpen
  \bibinfo {editor} {\bibfnamefont {G.}~\bibnamefont {Bertone}},\ ed.,\
  \href@noop {} {\emph {\bibinfo {title} {{Particle dark matter: Observations,
  models and searches}}}}\ (\bibinfo  {publisher} {Cambridge University
  Press},\ \bibinfo {year} {2010})\BibitemShut {NoStop}%
%%CITATION = INSPIRE-896695;%%
\bibitem [{\citenamefont {Read}\ \emph {et~al.}(2008)\citenamefont {Read},
  \citenamefont {Lake}, \citenamefont {Agertz},\ and\ \citenamefont
  {Debattista}}]{Read:2008fh}%
  \BibitemOpen
  \bibfield  {author} {\bibinfo {author} {\bibfnamefont {J.}~\bibnamefont
  {Read}}, \bibinfo {author} {\bibfnamefont {G.}~\bibnamefont {Lake}}, \bibinfo
  {author} {\bibfnamefont {O.}~\bibnamefont {Agertz}}, \ and\ \bibinfo {author}
  {\bibfnamefont {V.~P.}\ \bibnamefont {Debattista}},\ }\href@noop {}
  {\bibfield  {journal} {\bibinfo  {journal} {MNRAS}\ }\textbf {\bibinfo
  {volume} {389}},\ \bibinfo {pages} {1041} (\bibinfo {year} {2008})},\ \Eprint
  {http://arxiv.org/abs/0803.2714} {arXiv:0803.2714 [astro-ph]} \BibitemShut
  {NoStop}%
%%CITATION = ARXIV:0803.2714;%%
\bibitem [{\citenamefont {Bruch}\ \emph {et~al.}(2009)\citenamefont {Bruch},
  \citenamefont {Peter}, \citenamefont {Read}, \citenamefont {Baudis},\ and\
  \citenamefont {Lake}}]{Bruch:2009rp}%
  \BibitemOpen
  \bibfield  {author} {\bibinfo {author} {\bibfnamefont {T.}~\bibnamefont
  {Bruch}}, \bibinfo {author} {\bibfnamefont {A.~H.}\ \bibnamefont {Peter}},
  \bibinfo {author} {\bibfnamefont {J.}~\bibnamefont {Read}}, \bibinfo {author}
  {\bibfnamefont {L.}~\bibnamefont {Baudis}}, \ and\ \bibinfo {author}
  {\bibfnamefont {G.}~\bibnamefont {Lake}},\ }\href {\doibase
  10.1016/j.physletb.2009.03.042} {\bibfield  {journal} {\bibinfo  {journal}
  {Phys.Lett.}\ }\textbf {\bibinfo {volume} {B674}},\ \bibinfo {pages} {250}
  (\bibinfo {year} {2009})},\ \Eprint {http://arxiv.org/abs/0902.4001}
  {arXiv:0902.4001 [astro-ph.HE]} \BibitemShut {NoStop}%
\bibitem [{\citenamefont {Press}\ and\ \citenamefont
  {Spergel}(1985)}]{Press-Spergel}%
  \BibitemOpen
  \bibfield  {author} {\bibinfo {author} {\bibfnamefont {W.~H.}\ \bibnamefont
  {Press}}\ and\ \bibinfo {author} {\bibfnamefont {D.~N.}\ \bibnamefont
  {Spergel}},\ }\href {\doibase 10.1086/163485} {\bibfield  {journal} {\bibinfo
   {journal} {Astrophys.J.}\ }\textbf {\bibinfo {volume} {296}},\ \bibinfo
  {pages} {679} (\bibinfo {year} {1985})}\BibitemShut {NoStop}%
%%CITATION = ASJOA,296,679;%%
\bibitem [{\citenamefont {{Krauss}}\ \emph {et~al.}(1985)\citenamefont
  {{Krauss}}, \citenamefont {{Freese}}, \citenamefont {{Spergel}},\ and\
  \citenamefont {{Press}}}]{Krauss:1985ApJ}%
  \BibitemOpen
  \bibfield  {author} {\bibinfo {author} {\bibfnamefont {L.~M.}\ \bibnamefont
  {{Krauss}}}, \bibinfo {author} {\bibfnamefont {K.}~\bibnamefont {{Freese}}},
  \bibinfo {author} {\bibfnamefont {D.~N.}\ \bibnamefont {{Spergel}}}, \ and\
  \bibinfo {author} {\bibfnamefont {W.~H.}\ \bibnamefont {{Press}}},\ }\href
  {\doibase 10.1086/163767} {\bibfield  {journal} {\bibinfo  {journal} {\apj}\
  }\textbf {\bibinfo {volume} {299}},\ \bibinfo {pages} {1001} (\bibinfo {year}
  {1985})}\BibitemShut {NoStop}%
\bibitem [{\citenamefont {Freese}(1986)}]{Freese:1985qw}%
  \BibitemOpen
  \bibfield  {author} {\bibinfo {author} {\bibfnamefont {K.}~\bibnamefont
  {Freese}},\ }\href {\doibase 10.1016/0370-2693(86)90349-7} {\bibfield
  {journal} {\bibinfo  {journal} {Phys.Lett.}\ }\textbf {\bibinfo {volume}
  {B167}},\ \bibinfo {pages} {295} (\bibinfo {year} {1986})}\BibitemShut
  {NoStop}%
%%CITATION = PHLTA,B167,295;%%
\bibitem [{\citenamefont {Krauss}\ \emph {et~al.}(1986)\citenamefont {Krauss},
  \citenamefont {Srednicki},\ and\ \citenamefont {Wilczek}}]{Krauss:1985aaa}%
  \BibitemOpen
  \bibfield  {author} {\bibinfo {author} {\bibfnamefont {L.~M.}\ \bibnamefont
  {Krauss}}, \bibinfo {author} {\bibfnamefont {M.}~\bibnamefont {Srednicki}}, \
  and\ \bibinfo {author} {\bibfnamefont {F.}~\bibnamefont {Wilczek}},\ }\href
  {\doibase 10.1103/PhysRevD.33.2079} {\bibfield  {journal} {\bibinfo
  {journal} {Phys.Rev.}\ }\textbf {\bibinfo {volume} {D33}},\ \bibinfo {pages}
  {2079} (\bibinfo {year} {1986})}\BibitemShut {NoStop}%
%%CITATION = PHRVA,D33,2079;%%
\bibitem [{\citenamefont {Gaisser}\ \emph {et~al.}(1986)\citenamefont
  {Gaisser}, \citenamefont {Steigman},\ and\ \citenamefont
  {Tilav}}]{Gaisser:1986ha}%
  \BibitemOpen
  \bibfield  {author} {\bibinfo {author} {\bibfnamefont {T.}~\bibnamefont
  {Gaisser}}, \bibinfo {author} {\bibfnamefont {G.}~\bibnamefont {Steigman}}, \
  and\ \bibinfo {author} {\bibfnamefont {S.}~\bibnamefont {Tilav}},\ }\href
  {\doibase 10.1103/PhysRevD.34.2206} {\bibfield  {journal} {\bibinfo
  {journal} {Phys.Rev.}\ }\textbf {\bibinfo {volume} {D34}},\ \bibinfo {pages}
  {2206} (\bibinfo {year} {1986})}\BibitemShut {NoStop}%
%%CITATION = PHRVA,D34,2206;%%
\bibitem [{\citenamefont {Lundberg}\ and\ \citenamefont
  {Edsj{\"o}}(2004)}]{Lundberg}%
  \BibitemOpen
  \bibfield  {author} {\bibinfo {author} {\bibfnamefont {J.}~\bibnamefont
  {Lundberg}}\ and\ \bibinfo {author} {\bibfnamefont {J.}~\bibnamefont
  {Edsj{\"o}}},\ }\href {\doibase 10.1103/PhysRevD.69.123505} {\bibfield
  {journal} {\bibinfo  {journal} {Phys.Rev.}\ }\textbf {\bibinfo {volume}
  {D69}},\ \bibinfo {pages} {123505} (\bibinfo {year} {2004})},\ \Eprint
  {http://arxiv.org/abs/astro-ph/0401113} {arXiv:astro-ph/0401113 [astro-ph]}
  \BibitemShut {NoStop}%
\bibitem [{\citenamefont {Gould}(1987)}]{Gould:capture}%
  \BibitemOpen
  \bibfield  {author} {\bibinfo {author} {\bibfnamefont {A.}~\bibnamefont
  {Gould}},\ }\href {\doibase 10.1086/165653} {\bibfield  {journal} {\bibinfo
  {journal} {Astrophys.J.}\ }\textbf {\bibinfo {volume} {321}},\ \bibinfo
  {pages} {571} (\bibinfo {year} {1987})}\BibitemShut {NoStop}%
\bibitem [{\citenamefont {Gould}(1990)}]{Gould:diffusion}%
  \BibitemOpen
  \bibfield  {author} {\bibinfo {author} {\bibfnamefont {A.}~\bibnamefont
  {Gould}},\ }\href@noop {} {\bibfield  {journal} {\bibinfo  {journal}
  {Astrophys.J.}\ }\textbf {\bibinfo {volume} {368}},\ \bibinfo {pages} {610}
  (\bibinfo {year} {1990})}\BibitemShut {NoStop}%
\bibitem [{\citenamefont {Gould}\ and\ \citenamefont
  {Khairul~Alam}(2001)}]{GouldAlam}%
  \BibitemOpen
  \bibfield  {author} {\bibinfo {author} {\bibfnamefont {A.}~\bibnamefont
  {Gould}}\ and\ \bibinfo {author} {\bibfnamefont {S.}~\bibnamefont
  {Khairul~Alam}},\ }\href {\doibase 10.1086/319040} {\bibfield  {journal}
  {\bibinfo  {journal} {Astrophys.J.}\ }\textbf {\bibinfo {volume} {549}},\
  \bibinfo {pages} {72} (\bibinfo {year} {2001})},\ \Eprint
  {http://arxiv.org/abs/astro-ph/9911288} {arXiv:astro-ph/9911288 [astro-ph]}
  \BibitemShut {NoStop}%
%%CITATION = ASTRO-PH/9911288;%%
\bibitem [{\citenamefont {Damour}\ and\ \citenamefont {Krauss}(1999)}]{Damour}%
  \BibitemOpen
  \bibfield  {author} {\bibinfo {author} {\bibfnamefont {T.}~\bibnamefont
  {Damour}}\ and\ \bibinfo {author} {\bibfnamefont {L.~M.}\ \bibnamefont
  {Krauss}},\ }\href {\doibase 10.1103/PhysRevD.59.063509} {\bibfield
  {journal} {\bibinfo  {journal} {Phys.Rev.}\ }\textbf {\bibinfo {volume}
  {D59}},\ \bibinfo {pages} {063509} (\bibinfo {year} {1999})},\ \Eprint
  {http://arxiv.org/abs/astro-ph/9807099} {arXiv:astro-ph/9807099 [astro-ph]}
  \BibitemShut {NoStop}%
\bibitem [{\citenamefont {Peter}(2009{\natexlab{a}})}]{Peter:first}%
  \BibitemOpen
  \bibfield  {author} {\bibinfo {author} {\bibfnamefont {A.~H.}\ \bibnamefont
  {Peter}},\ }\href {\doibase 10.1103/PhysRevD.79.103531} {\bibfield  {journal}
  {\bibinfo  {journal} {Phys.Rev.}\ }\textbf {\bibinfo {volume} {D79}},\
  \bibinfo {pages} {103531} (\bibinfo {year} {2009}{\natexlab{a}})},\ \Eprint
  {http://arxiv.org/abs/0902.1344} {arXiv:0902.1344 [astro-ph.HE]} \BibitemShut
  {NoStop}%
%%CITATION = ARXIV:0902.1344;%%
\bibitem [{\citenamefont {Peter}(2009{\natexlab{b}})}]{Peter:sun}%
  \BibitemOpen
  \bibfield  {author} {\bibinfo {author} {\bibfnamefont {A.~H.}\ \bibnamefont
  {Peter}},\ }\href {\doibase 10.1103/PhysRevD.79.103532} {\bibfield  {journal}
  {\bibinfo  {journal} {Phys.Rev.}\ }\textbf {\bibinfo {volume} {D79}},\
  \bibinfo {pages} {103532} (\bibinfo {year} {2009}{\natexlab{b}})},\ \Eprint
  {http://arxiv.org/abs/0902.1347} {arXiv:0902.1347 [astro-ph.HE]} \BibitemShut
  {NoStop}%
\bibitem [{\citenamefont {Sivertsson}\ and\ \citenamefont
  {Edsj{\"o}}(2010)}]{Sivertsson}%
  \BibitemOpen
  \bibfield  {author} {\bibinfo {author} {\bibfnamefont {S.}~\bibnamefont
  {Sivertsson}}\ and\ \bibinfo {author} {\bibfnamefont {J.}~\bibnamefont
  {Edsj{\"o}}},\ }\href {\doibase 10.1103/PhysRevD.81.063502} {\bibfield
  {journal} {\bibinfo  {journal} {Phys.Rev.}\ }\textbf {\bibinfo {volume}
  {D81}},\ \bibinfo {pages} {063502} (\bibinfo {year} {2010})},\ \Eprint
  {http://arxiv.org/abs/0910.0017} {arXiv:0910.0017 [astro-ph.HE]} \BibitemShut
  {NoStop}%
\bibitem [{\citenamefont {Peter}(2009{\natexlab{c}})}]{Peter:earth}%
  \BibitemOpen
  \bibfield  {author} {\bibinfo {author} {\bibfnamefont {A.~H.}\ \bibnamefont
  {Peter}},\ }\href {\doibase 10.1103/PhysRevD.79.103533} {\bibfield  {journal}
  {\bibinfo  {journal} {Phys.Rev.}\ }\textbf {\bibinfo {volume} {D79}},\
  \bibinfo {pages} {103533} (\bibinfo {year} {2009}{\natexlab{c}})},\ \Eprint
  {http://arxiv.org/abs/0902.1348} {arXiv:0902.1348 [astro-ph.HE]} \BibitemShut
  {NoStop}%
\end{thebibliography}%

\end{document}